\pdfoutput=1

\documentclass[12pt,a4paper,hyperref]{JHEP3}
\let\ifpdf\relax
\usepackage{ifpdf}
\usepackage{color}
\let\normalcolor\relax
\usepackage{mathtools}
\usepackage{cite}
\usepackage{graphicx,subcaption} 

\renewcommand{\hat}{\widehat}
\renewcommand{\tilde}{\widetilde}
\renewcommand{\phi}{\varphi}
\DeclareMathOperator*{\Res}{\mathrm{Res}}
\newcommand{\chyInt}{d\Omega_{\mathrm{CHY}}}
\newcommand{\mi}{\raisebox{0.75pt}{\scalebox{0.75}{$\,-\,$}}}
\newcommand{\pl}{\raisebox{0.75pt}{\scalebox{0.75}{$\,+\,$}}}

\newcommand{\fwbox}[2]{\text{\makebox[#1][c]{$\hspace{-150pt}\displaystyle#2\hspace{-150pt}$}}}

\newcommand{\fwboxR}[2]{\text{\makebox[#1][r]{$#2$}}}
\newcommand{\eq}[1]{\vspace{-2.5pt}\begin{equation}\hspace{-100pt}#1\hspace{-100pt}\vspace{-2.5pt}\end{equation}}
\newcommand{\eqs}[1]{\vspace{-0.0pt}\begin{equation}\begin{split}#1\end{split}\vspace{-0.0pt}\end{equation}}
\newcommand{\fig}[3]{\raisebox{#1}{\includegraphics[scale=#2]{#3}}}
\newcommand{\Pfprime}{\text{Pf}\,'\!\hspace{1pt}}
\newcommand{\z}[2]{(z_{#1}\!-\!z_{#2})}

\preprint{2015}
\title{\mbox{{\LARGE Integration Rules for Scattering Equations}}}
\author{{\normalsize Christian~Baadsgaard, N.~E.~J.~Bjerrum-Bohr, Jacob~L.~Bourjaily, and Poul~H.~Damgaard}\\
\mbox{{Niels Bohr International Academy and Discovery Center,}}\\
\mbox{{Niels Bohr Institute, University of Copenhagen,}}\\
\mbox{Blegdamsvej 17, DK-2100 Copenhagen \O, Denmark}\vspace{-8pt}}
\abstract{
As described by Cachazo, He and Yuan, scattering amplitudes in many quantum field theories can be represented 
as integrals that are fully localized on solutions to the so-called scattering equations. Because the number of solutions 
to the scattering equations grows quite rapidly, the contour of integration involves contributions from many 
isolated components. In this paper, we provide a simple, combinatorial rule that immediately provides the 
result of integration against the scattering equation constraints for any M\"{o}bius-invariant integrand 
involving only simple poles. These rules have a simple diagrammatic interpretation that makes the evaluation of 
any such integrand immediate. Finally, we explain how these rules are related to the computation of 
amplitudes in the field theory limit of string theory.}

\begin{document}\newpage
\section{Introduction}\vspace{-5pt}
A most surprising new way of computing field theory amplitudes discovered by Cachazo,
He and Yuan (CHY)~\cite{Cachazo:2013gna,Cachazo:2013hca,Cachazo:2013iea} uses as essential input the 
solutions to a set of so-called scattering equations, which depend on the external momenta. Amplitudes can 
then be represented as integrals over auxiliary variables that are fully localized by constraints from the 
scattering equations. This means that tree-level field theory amplitudes of an arbitrary number of external 
legs can be computed by solving a set of algebraic equations, and summing over contributions from these 
solutions. Early on, it was shown how the construction extends from scalar $\phi^3$-theory
to Yang-Mills and gravity~\cite{Cachazo:2013hca}. And a direct proof of the general construction
for scalar $\phi^3$-theory and Yang-Mills theory has been provided by Dolan and
Goddard~\cite{Dolan:2013isa}.

Much in the construction of the CHY formalism resembles string theory. Indeed,
the CHY prescription can be seen as a particular infinite-tension limit of
ambitwistor strings~\cite{Mason:2013sva, Berkovits:2013xba, Gomez:2013wza,
Adamo:2013tsa, Ohmori:2015sha}.
More generally, it was shown in \mbox{ref.\ \cite{Bjerrum-Bohr:2014qwa}} that 
ordinary superstring theory can be put in a form that immediately transcribes into
a CHY-type construction by a simple change of integration measure. In this way, 
amplitudes involving fermions, scalars, and combinations thereof with gauge fields
could be given a CHY-like prescription, directly from superstring integrands. (For an 
alternative discussion of amplitudes with fermions, see also \mbox{ref.~\cite{Weinzierl:2014ava}.})
This transcription was shown to be possible once all tachyonic modes had
been made to cancel explicitly in the integrand by means of integrations by parts.
In this way, a very large class of theories have been given a CHY-like prescription.
Extensions to include massive legs~\cite{Naculich:2014naa,Naculich:2015zha} have also been considered. 
A connection between kinematic algebra for gauge theory and the scattering equations have been
considered in \mbox{ref.\ \cite{Monteiro:2013rya}}.

More recently, Cachazo, He and Yuan have generalized the construction to
$\phi^4$-theory and various related scalar theories coupled to gauge fields and 
gravitons~\cite{Cachazo:2014xea}. Although based on dimensional reduction of Yang-Mills theory, this
provides yet more evidence that a CHY-like prescription may exist for any
quantum field theory. 

Essential to the CHY formulation is the need to solve the constraints that impose the scattering 
equations. For an $n$-particle amplitude, there are $(n-3)!$ such solutions. This is obviously related to the fact that 
BCJ-identities~\cite{Bern:2008qj} reduce the basis of $n$-particle amplitudes
to one involving only $(n-3)!$ partial amplitudes~\cite{BjerrumBohr:2009rd, Stieberger:2009hq}. 
While straightforward in principle, the need to find $(n-3)!$ solutions to the scattering equations
rapidly becomes a (serious) computational obstacle. Moreover, after summing over all the different solutions, 
one often ends up with a remarkably simple answer. Such a situation is not uncommon in
quantum field theory, and it is natural to wonder if there were a way around it. Is there a simpler 
way to produce the result obtained after summing over all solutions to the scattering equations? 

The purpose of this paper is to arrive at such direct integration rules. String theory can provide a very useful 
source of inspiration for this problem. By considering integrands with simple poles, tree-level field theory amplitudes can be obtained as a singular
limit near the poles, with explicit factors of $\alpha'$ outside the integral 
rendering the final answer finite as $\alpha'\!\!\to\! 0$. With the explicit map provided
in \mbox{ref.\ \cite{Bjerrum-Bohr:2014qwa}} from superstring integrands to CHY integrands, it is clear 
that a relation must exist between these two very different types of integrals. One of
the aims of this paper is to derive this relationship explicitly.

The class of integrands to which the integration rules apply are $\text{SL}(2,{\mathbb C})$-invariant 
CHY integrands that have simple poles only, but which can have non-trivial
numerators. These integrands are special in that their integrals evaluate 
to a sum of Feynman diagrams---the most interesting class of integrands. 
It has previously been shown by CHY~\cite{Cachazo:2013hca} that a certain 
class of integrals produce sums of Feynman diagrams. This paper generalizes this result and characterizes 
all CHY integrals that are equal to Feynman diagrams. In addition, we provide a set of integration rules for 
evaluating these integrals. The integration rules can be given in a precise algebraic form, but can also 
be conveniently described diagrammatically. Using our diagrammatic method, one can essentially read-off 
the result of integration with respect to the CHY measure and the associated sum over $(n-3)!$ solutions.

In order to calculate Yang-Mills or gravity amplitudes in the CHY formalism, it is necessary to also 
consider CHY integrals with higher order poles. We therefore provide a procedure for reducing such 
integrals into those that can be evaluated with the integration rules we have derived. This link will be provided by a set of Pfaffian identities that are valid 
on solutions to the scattering equations. This procedure is easy to execute---though it provides an alternative to the method recently described in \mbox{ref.\ \cite{Cachazo:2015nwa}} for evaluating CHY integrals.

We also give integration rules for the CHY integrands that contain a Pfaffian and appear in the CHY 
$\phi^4$-theory~\cite{Cachazo:2014xea}, and we examine the connection to the dual formulation in string theory. 

\newpage
\section{From Superstring Amplitudes to CHY integrands}\vspace{-5pt}

In order to develop integration rules for CHY integrands it is useful to first describe the 
link between superstring theory and the CHY prescription~\cite{Bjerrum-Bohr:2014qwa}. 
The CHY representation for the $n$-particle amplitude in $\phi^3$-theory can be given 
as,~\cite{Cachazo:2013gna,Cachazo:2013hca,Cachazo:2013iea,Dolan:2013isa}:
\eq{\mathcal{A}_n^{\phi^3}=g^{n-2}\int\!\chyInt\,\,\left(\frac{1}{\z{1}{2}^2\z{2}{3}^2\!\cdots\z{n}{1}^2}\right),\label{scalarCHY}}
where $\chyInt$ denotes the following integration measure combined with the 
scattering equation constraints,\\
\eq{\hspace{-0.8cm}\chyInt\equiv\frac{d^nz}{\mathrm{vol}(\text{SL}(2,\mathbb{C})\!)}
\prod_i\,\!'\delta\big(S_i\big)=\z{r}{s}^2\z{s}{t}^2\z{t}{r}^2
\prod_{\fwbox{25pt}{i\!\in\!\mathbb{Z}_n\!\backslash\{r,s,t\}}}dz_i\,\delta\big(S_i\big)\,,\label{definition_of_chy_measure}}
(independent of the choice of $\{r,s,t\}$), where $S_i$ denotes the $i^{\mathrm{th}}$ scattering equation,
\eq{S_i\equiv\sum_{j\neq i}\frac{s_{ij}}{\z{i}{j}}\equiv\sum_{j\neq i}\frac{2 k_i\!\cdot\!k_j}{\z{i}{j}}\,,}
with $k_i$ and $k_j$ on-shell, massless momenta so that $s_{ij}\!\equiv\!(k_i+k_j)^2=2k_i\!\cdot\!k_j$. 
In general, we define $s_{i,j,\cdots,k}\!\equiv\!(k_i\pl k_j\pl\cdots\pl k_k)^2$. As usual, the $\delta$-functions are to be understood 
homomorphically as residue prescriptions:
\eq{\int\!dz\,h(z)\delta\big(f(z)\big)\equiv\Res_{f(z)=0}(dz\,h(z)/f(z))\,.}

Compare this with a string-like representation of a corresponding amplitude,
\eq{\hspace{-20pt}{\cal A}_n \!= \lim_{\alpha'\!\to 0}g^{n-2}\,{\alpha'\hspace{1pt}}^{n-3}\,
\int\prod_{i=3}^{n-1}
dz_i\,{(z_1 - z_2)(z_2-z_n)(z_n - z_1)\over \prod_{i=1}^n (z_i - z_{i+1})}
\prod_{1\leq i< j\leq n} |z_i-z_j|^{\alpha' s_{ij}}\!
\,, \label{scalardual}}
one sees immediately that the CHY prescription (\ref{scalarCHY}) is obtained by
substituting the $\delta$-function constraint
\eq{{\alpha'}^{3-n} (z_1-z_2)(z_2-z_n) (z_n-z_1)\prod_{i\neq1,2,n}\delta(S_i)
\prod_{i=1}^{n} (z_i - z_{i+1})^{-1}\,,
\label{CHYstringlink}}
into the string-theory measure.\footnote{Here, we have chosen the values $\{1,2,n\}$ for $\{r,s,t\}$ 
in~\eqref{definition_of_chy_measure}.} This is indeed the general prescription for
the superstring---hence for gauge fields, scalars and fermions---once all tachyon poles have been explicitly cancelled through
partial integrations~\cite{Bjerrum-Bohr:2014qwa}. Below, we will also discuss cases
where a match can be made between bosonic string theory and CHY integrands.

It is instructive to analyze how the two expressions, (\ref{scalarCHY}) and 
(\ref{scalardual}), can coincide. First, in string theory, the integration contour is explicitly ordered. This directly encodes the cyclic ordering of
legs in the amplitude. In contrast, the integration contour for (\ref{scalarCHY}) is not explicitly ordered; rather, the cyclic ordering is encoded in terms of the poles in the measure of integration. Second, the regions of integrations that contribute in the $\alpha'\!\!\to\! 0$ limit of
(\ref{scalardual}) are localized around the singularities of the
poles, while the CHY measure hits precisely the correct {\em fixed} values that 
reproduce the leading result of the integral.

Pure Yang-Mills theory is only slightly more complicated. In the superstring formalism
reviewed in \mbox{ref.\ \cite{Green:1987sp}}, an $n$-point field theory amplitude can 
be computed through ordered integrations as follows:
\eqs{\hspace{-20pt}\mathcal{A}_n=&\lim_{\alpha'\!\to0}{\alpha'}^{(n-4)/2}\,
\int\prod_{i=3}^{n-1}
dz_i\,{(z_1\mi z_{2})(z_{2}\mi z_n)(z_n\mi z_1)\over \prod_{i=1}^n (z_i - z_{i+1})}\\
&\times\int\prod_id\theta_i\prod_jd\phi_j\prod_{i<j}(z_i-z_j-\theta_i\theta_j)^{\alpha's_{ij}}\\
&\times\prod_{i<j}\exp\left[\frac{\sqrt{2\alpha'}(\theta_i\mi\theta_j)
(\phi_i\varepsilon_i\!\cdot\!k_j\pl\phi_j\varepsilon_j\!\cdot\!k_i)}
{z_i-z_j} \right.\mi\left. \frac{\phi_i\phi_j\varepsilon_i\!\cdot\!\varepsilon_j}{z_i-z_j}\mi\frac{\theta_i\theta_j\phi_i\phi_j\varepsilon_i\cdot\varepsilon_j}{(z_i-z_j)^2}\right]\!,\hspace{-30pt}}
where the auxiliary, Grassmann integrations over $\phi_i$ and $\theta_i$ automatically
impose the multi-linearity condition on the amplitude in terms of external polarization
vectors $\varepsilon_j^{\mu}$. The result of performing these Grassmann integrations will be ordinary bosonic integrands that have poles in the $z_i$ variables. By repeated use of integrations by parts,
or alternatively, by performing the superstring computation with the picture change of
\mbox{ref.\ \cite{Bjerrum-Bohr:2014qwa}}, this bosonic integrand can be written in terms of single poles only
plus terms that are proportional to the scattering equations. This separation of terms
is equivalent to a complete cancellation of tachyon poles in the superstring integrand.
At this point, one can insert the $\delta$-function constraint (\ref{CHYstringlink}) to recover
the CHY prescription~\cite{Cachazo:2013gna,Cachazo:2013hca,Cachazo:2013iea} for Yang-Mills theory. 

This connection between a string theory computation of field theory amplitudes and the
CHY formalism leads us to investigate in detail the manner in which the two different
integrands produce the same answer. To make the discussion general, let us consider the
following limit of a generic ordered string theory integral,
\eq{\hspace{-20pt}{\cal I}_n \!= \lim_{\alpha'\!\to 0}{\alpha'\hspace{1pt}}^{n-3}\,
\int\prod_{i=3}^{n-1}
\!\!dz_i\,(z_1\mi z_{2})(z_{2}\mi z_n)(z_n\mi z_1) 
\prod_{\!\!1\leq i< j\leq n\!\!} |z_i-z_j|^{\alpha' s_{ij}}H(z)\,,}
where $H(z)$ consists of products of factors $(z_i - z_j)^{-\ell}$ such that the whole integrand is 
M\"{o}bius-invariant. It is often convenient to consider $H(z)$ prior to gauge-fixing.

We are mostly interested in string theory integrands $H(z)$ that can lead to convergent integrals
for finite $\alpha'$ in a neighborhood around the origin. 
As mentioned above, this can be achieved in the superstring case by combining terms, perhaps
after integration by parts. With the prefactor $\alpha'\hspace{1pt}^{n-3}$ this means that to obtain a finite
field theory limit (at tree-level), the leading divergence will go as $1/\alpha'\hspace{1pt}^{n-3}$. The
$n-3$ integrations remaining after gauge-fixing provide this leading term. Let us now explore
in detail how this $1/\alpha'\hspace{1pt}^{n-3}$-divergence is achieved and find the finite result it leaves
as $\alpha'\!\!\to\! 0$, after cancellation with the prefactor. We may determine whether or not the
integral will diverge with the needed power as follows:

\vspace{0.2cm}
\noindent
{\em The integration rule:}\\
\indent\textbullet\,\,Enumerate all subsets of consecutive variables $T\!\equiv\!\{z_j,z_{j+1},\ldots,z_{j+m}\}$ for which the number of factors $(z_l-z_k)^{-1}$ in $H(z)$ (with multiplicity) for pairs $\{z_l,z_k\}\!\in\!T$ is equal to $m$---the number of elements in $T$ minus one. Complementary subsets are to be taken to be equivalent, $T\!\simeq\!T^c$. Integration over the 
variables of each such subsets has a $1/\alpha'$-divergence in the $\alpha'\!\!\to\!0$ limit. To each subset $T$, we assign a factor $1/(\alpha's_{j,j+1,\ldots,j+m})$.

\indent\textbullet\,\,Let us call a pair of enumerated subsets $T_1,T_2$ satisfying the criterion above {\it compatible} if they (or their complements, which are considered equivalent) are either nested or disjoint.

\indent\textbullet\,\,Every $(n\mi3)$-element collection $\tau\!\equiv\!\{T_1,\ldots,T_{n-3}\}$ of pairwise-compatible enumerated subsets will contribute the product of the factors for each $T_a\!\in\!\tau$ to the integral. The final integral is simply the sum over these products for each of the collections $\tau$.

It follows from these rules that if there are no $(n\mi3)$-element collections of mutually compatible subsets, the integral (combined with the $\alpha'\hspace{1pt}^{n-3}$ prefactor) will vanish in the $\alpha'\!\!\to\!0$ limit. 

\vspace{0.2cm}
\noindent
{\em Sketch of proof:}\\ 
Consider the integral,
\eq{\hspace{-20pt}{\cal I}_n \!= \lim_{\alpha'\!\to 0}{\alpha'\hspace{1pt}}^{n-3}\,
\int\prod_{i=3}^{n-1}
dz_i\,(z_1\mi z_{2})(z_{2}\mi z_n)(z_n\mi z_1) 
\prod_{\!\!1\leq i< j\leq n\!\!} |z_i-z_j|^{\alpha' s_{ij}} H(z)\,.}
The integration rules are formulated without reference to a specific gauge, but to derive them it is convenience to work in a specific gauge. We choose the gauge where $z_1\!=\!\infty$, $z_2\!=\!1$, $z_n\!=\!0$. 
The integrations are then ordered,
\eq{\int_0^1 \!\!dz_3\int_0^{z_3}\!\!dz_4\cdots\int_0^{z_{n-2}}\!\!dz_{n-1}\,.}

A divergence as $\alpha'\!\!\to\! 0$ occurs when some of the variables $z_i$, $i\!\in\!\{2,\ldots,n\}$ 
tend to the same value. Because of the ordered integration domain, these variables must be consecutive.
Now pick a positive integer $m$ and consider the case when variables $z_j$ to $z_{j+m}$ tend 
to the same value. It is useful to define $y_j \!\equiv\! 0$ and to introduce new variables 
for $i\!=\!j+1,\ldots,j+m$:
\eq{y_{i} \equiv(z_j-z_{i})\,.}
Then, for $k,l\!\in\!\{j,j+1,\ldots,j+m\}$\,,\vspace{5pt}
\eq{(z_k-z_l) \to (y_l-y_k)\,.}
and the integration region becomes
\eq{\hspace{-20pt}\int_0^{z_j} dz_{j+1}\cdots\int_0^{z_{j+m-1}} dz_{j+m} = 
\int_0^{z_j} dy_{j+m}\int_0^{y_{j+m}} dy_{j+m-1}\cdots\int_0^{y_{j+2}} dy_{j+1}\,.}
In order to identify potential divergences in the $\alpha'\!\!\to\! 0$ limit, it is convenient
to define $x_{j+m}\!\equiv\!1$ and perform one more change of variables for $i\!=\!j+1,\ldots,j+m-1$:
\eq{x_i \equiv y_i/y_{j+m} \,,}
so that, for $k,l\!\in\!\{j,j+1,j+2,\ldots,j+n\}$,
\eq{(y_l-y_k) \to (y_{j+m})(x_l-x_k) \,.}
The Jacobian of this change of variables is therefore $(y_{j+m})^{m-1}$, and the integration is now,
\eqs{
&\int_0^{z_j} dy_{j+m}\int_0^{y_{j+m}} dy_{j+m-1}\cdots\int_0^{y_{j+2}} dy_{j+1}=\\
&\int_0^{z_j} dy_{j+m}(y_{j+m})^{m-1}\int_0^{1} dx_{j+m-1}\int_0^{x_{j+m-1}} dx_{j+m-2}\cdots\int_0^{x_{j+2}} dx_{j+1}\,.}
The advantage of this transformation is that the integration domains of variables $x_{i}$ 
do not depend on $y_{j+m}$. Whether or not the integral over $y_{j+m}$ will diverge at
the lower endpoint 
will therefore only depend on whether the integrand contains $y_{j+m}$ raised to any negative power.

For each pole $(z_l-z_k)^{-1}$ in $H(z)$ with $k,l\!\in\!\{j,j+1,j+2,\ldots,j+m\}$ we pick up 
a factor of $(y_{j+m})^{-1}$. Consequently, if the number of such factors is equal to $m$, 
we get a factor of $(y_{j+m})^{-m}$. After including the Jacobian, this leaves us with
$(y_{j+m})^{-1}$. That is, when $H(z)$ has precisely $m$ such factors (counted with multiplicity),
we obtain a divergence that goes as $1/\alpha'$ as $\alpha'\!\!\to\! 0$.\footnote{ 
If we consider the more general case where there can also be factors in the numerator of $H(z)$, 
the argument goes through essentially unchanged. It is then the number of factors going to zero 
in the denominator minus those in the numerator that must equal $m$.} And because of the factor,
\eq{\prod_{1\leq i< j\leq n} |z_i-z_j|^{\alpha' s_{ij}},\label{knfactor}}
in the integration measure, we also get a factor of $(y_{j+m})^{\alpha's_{kl}}$ for each pair 
$k,l\!\in\!\{j,j+1,j+2,\ldots,j+m\}$. In other words, under the two changes of variables described above, 
\eqref{knfactor} produces a factor of $(y_{j+m})^{\alpha's_{j,j+1,\ldots,j+m}}$. 
We therefore arrive at the following integral:
\eq{\int_0^{z_j} dy_{j+m} \left(\frac{(y_{j+m})^{\alpha's_{j,j+1,\ldots,j+m}}}{y_{j+m}} + \cdots \right) =
\frac{1}{\alpha's_{j,j+1,\ldots,j+m}}+\mathcal{O}\big(({\alpha'})^{0}\big)\,,}
where the terms in the ellipsis are less singular in the $\alpha'\!\!\to\! 0$ limit.

The final task is to collect all such leading divergences in the limit $\alpha'\!\!\to\! 0$.
Retaining only these leading-order terms, we can neglect all factors of the 
form $(z_l-z_k)$ when $l\!\in\!\{j,j+1,\ldots,j+m\}$ while $k \not\in \{j,j+1,\ldots,j+m\}$ (or vice versa). 
The full string theory integral therefore factors into three parts: (i) the integrals over $z_i$ 
for $i\!<\!j$; (ii) the integrals over $z_i$ for $j\!\leq\! i \!\leq\! j+m$; and (iii) the integrals 
over $z_i$ for $j + m \!<\! i$. Each of these three parts will be of the same form as the full original integral.
It is evident that all factors $1/s_{i_1,i_2,\ldots, i_s}$ so obtained from these contributions will be either
{\em disjoint} or {\em nested}. But it can happen that integrations in overlapping sets each have the required divergence. This occurs when variables $z_{i}$ to $z_{j}$ tend to the same value, but where we
also get a divergence when variables $z_{k}$ to $z_{l}$ tend to the same value, with the ordering 
$i \!<\! k \!<\! j \!<\! l$. In this case the divergences occur in two distinct regions of the integration 
domain of the full integral. Since we can write the full integral as the sum of the integral 
over one part of the domain plus the integral over the remaining part, we must sum 
over the contributions from the two divergent regions. In general we must sum over all the 
contributions from distinct regions with the required divergence. It is perhaps easiest to illustrate the complete integration rule by a simple example.

\vspace{0.2cm}
\noindent
{\em Example:}\\
Consider the string theory integral that involves the following $H(z)$ factor:
\eq{H(z) = \frac{1}{(z_1-z_2)(z_1-z_5)(z_2-z_4)(z_3-z_4)(z_3-z_6)(z_5-z_6)} \,.}
To evaluate the integral, we first use the rule saying that complementary subsets are equivalent and 
select from the pairs of equivalent subsets the one not containing $z_1$. 
Then the subsets of the variables that will yield a $1/\alpha'$-divergence are the following:\\[-12pt]
\eqs{& \{3,4\}: \text{ two variables, one factor connecting them} \\
& \{5,6\}: \text{ two variables, one factor connecting them} \\
& \{2,3,4\}: \text{ three variables, two factors connecting them} \\
& \{3,4,5,6\}: \text{ four variables, three factors connecting them}}
These subsets are all compatible with each other except that $\{2,3,4\}$ and 
$\{3,4,5,6\}$ are incompatible. We can therefore form two collections of three, pairwise compatible subsets:
\eqs{
&\tau_1\equiv\big\{\!\{3,4\},\{5,6\},\{2,3,4\}\!\big\}\\
&\tau_2\equiv\big\{\!\{3,4\},\{5,6\},\{3,4,5,6\}\!\big\}}%
Consequently, using our integration rule, we find that to leading-order in $\alpha'$ the 
integral is given as follows:
\eq{\frac{1}{s_{34}s_{56}}\left(\frac{1}{s_{234}}+\frac{1}{s_{3456}}\right)\frac{1}{(\alpha')^3}\,.}
With the prefactor $(\alpha')^{6-3} = (\alpha')^3$ the leading contribution to this
string theory integral is therefore
\eq{\frac{1}{s_{34}s_{56}}\left(\frac{1}{s_{234}}+\frac{1}{s_{3456}}\right) \,.}
We note that this corresponds to the sum of two Feynman diagrams contributing to the 
6-point amplitude of $\phi^3$-theory.

It is easy to extend the above integration rules and derivation to the case when $H(z)$ also contains 
factors of $(z_i-z_j)$. The only change is that when considering a subset 
$T\!=\!\{z_j,z_{j+1},\ldots,z_{j+m}\}$, the condition for the integration over the variables in $T$ to have a 
$1/\alpha'$-divergence becomes the following: the number of factors $(z_l-z_k)^{-1}$ in $H(z)$ with $z_k,z_l\!\in\!T$ \emph{minus the number of factors} 
$(z_q-z_r)$ \emph{in} $H(z)$ \emph{with} $z_q,z_r\!\in\!T$ must equal $m$.

\vspace{-10pt}
\section{Rules for CHY Integration: the Global Residue Theorem} \label{GRT}\vspace{-5pt}

With the translation rule between string theory and CHY integrals of~\cite{Bjerrum-Bohr:2014qwa}, 
we know that the integration rules for the leading terms of string theory integrals
should map directly to integration rules for CHY integrals. The purpose of this
subsection is to establish that link directly. Our tool is the global residue theorem,
already exploited in detail by Dolan and Goddard in \mbox{ref.\ \cite{Dolan:2013isa}}. In this
section, we will first describe how to derive general integration rules for CHY
integrals. Then as a special case, we will limit ourselves to those kinds of integrals
(using the transcription prescription of \mbox{ref.\ \cite{Bjerrum-Bohr:2014qwa}}) that
were evaluated in the field theory limit of string theory in the previous section,
and demonstrate that they agree. This explicitly ties together the CHY prescription in terms
of the global residue theorem with the field theory limit of string theory, demonstrating
the equivalence of the two.

Using the transcription rule of \mbox{ref.\ \cite{Bjerrum-Bohr:2014qwa}} and comparing with the previous
section, we are led to consider CHY integrals of the following form:
\eq{I_n=\int \!\!\chyInt\,\,H(z)\,,}
where $\chyInt$ was defined in (\ref{definition_of_chy_measure}), and $H(z)$ is a product of factors $(z_i-z_j)^{-1}$ such that each $z_i$, with 
$i\!\in\!\{1,2,\ldots,n\}$, appears in exactly four terms (this ensures $\text{SL}(2,{\mathbb C})$-invariance).
Here, the integration contour is entirely localized by the scattering equation $\delta$-function constraints appearing in $\chyInt$. Interpreted as a contour integral (really, the residue) enclosing the solutions to the scattering equations,
\eq{\hspace{-20pt}I_n =\oint\displaylimits_{\!\!\!S_1=\cdots=S_n=0\!\!\!} \!\!\!\!\!dz_1dz_2\cdots dz_n \frac{(z_r-z_s)^2(z_s-z_t)^2(z_t-z_r)^2}
{dz_rdz_sdz_t}H(z)\frac{1}{\prod_{i\neq r,s,t} S_{i}}\,,}
enables us to make use of the global residue theorem. This is nicely explained and pursued
in \mbox{ref.\ \cite{Dolan:2013isa}}, and it tells us that the sum of the residues at the solutions to the scattering equations is equal 
to minus the sum of all other residues. By identifying and evaluating these other residues, 
we can therefore evaluate CHY integrals.

For concreteness, we will adopt the same gauge as in the previous section: \mbox{$z_1\!=\!\infty$}, $z_2\!=\!1$, $z_n\!=\!0$.
We also introduce, 
\eq{G(z_3,z_4,\ldots,z_{n-1}) ~\equiv~ \lim_{z_1\rightarrow \infty} \frac{z_1^4}{H(z_1,1,z_3,\ldots,z_{n-1},0)}\,.}
In this gauge, the CHY integral reads as follows:
\eq{I_n =\oint\displaylimits_{\!\!\!S_1=\cdots=S_n=0\!\!\!}  \!\!\!\!\!dz_3dz_4\cdots dz_{n-1} 
\frac{1}{G(z_3,z_4,\ldots,z_{n-1})}\frac{1}{\prod_{i=3}^{n-1} S_{i}}\,.}
It is convenient to have the integrand explicitly written out as a rational function. 
To this end, following Dolan and Goddard~\cite{Dolan:2013isa}, we introduce the following polynomial:
\eq{f_i = S_i \prod_{\substack{j=2\\j\neq i}}^n(z_i-z_j)\,.}
With these definitions we arrive at the following expression for the CHY integral:
\eq{I_n =\oint\displaylimits_{\!\!\!f_3=\cdots=f_{n-1}=0\!\!\!}  \!\!\!\!\!dz_3dz_4\cdots dz_{n-1} 
\frac{1}{G(z_3,z_4,\ldots,z_{n-1})}\frac{\prod_{i=3}^{n-1}\prod_{\substack{j=2\\j\neq i}}^n(z_i-z_j)}{\prod_{i=3}^{n-1} f_{i}}\,.}
Clearly, the task is to determine which conditions $G(z_3,z_4,\ldots,z_{n-1})$ must satisfy in order for 
the integrand of $I_n$ to contain poles beyond those at the solutions to the scattering equations. 

We now start to analyze this contour integral. Consider the case where variables $z_i$ tend to $z_n=0$
for $i\!\in\!T\!\equiv\!\{n-m,n-m+1,\ldots,n-1,n\}$ and $1\!\leq\! m\!\leq\! n-3$. Let us redefine:
\begin{align*}
z_{n-m} = \epsilon, \hspace{9mm} z_i = \epsilon \hspace{0.5mm} x_i \hspace{5mm} 
\text{for} \hspace{5mm} i=n-m,\hspace{1mm}n-m+1,\hspace{1mm}\ldots,\hspace{1mm}n\,.
\end{align*}
Note that $x_{n-m}=1$ and $x_n=0$.

We can determine whether the integrand $I_n$ has a pole at $\epsilon=0$ by simple power counting.
First, we count the powers of $\epsilon$ in the numerator: the factor 
\eq{\prod_{i=3}^{n-1}\prod_{\substack{j=2\\j\neq i}}^n(z_i-z_j)}
contains $m\!\times\!m$ factors of the form $(z_i-z_j)$ with $i,j\!\in\!T$ and each of these lead 
to one power of $\epsilon$. The measure $dz_3dz_4\cdots dz_{n-1}$ contributes $m-1$ factors of $\epsilon$, 
namely one factor for each $dz_i$ with $i\!\in\!T\backslash \{n-m,n\}$. 
So there are $m^2+m-1$ factors of $\epsilon$ in the numerator. 

In the denominator, each polynomial $f_i$ with $i\!\in\!T$ contains $m-1$ factors of $\epsilon$, 
and there are $m$ such $f_i$'s. If we denote the number of factors of $\epsilon$ in $G(z_3,z_4,\ldots,z_{n-1})$ 
with $m_G$, there are $m\!\times\!(m-1)+m_G$ factors of $\epsilon$ in the denominator. 
Subtracting the $\epsilon$-factors in the numerator, we find that there are $m_G-2m+1$ net factors 
of $\epsilon$ in the denominator. In conclusion, we arrive at the counting rule that if $m_G\!=\!2m$, 
then there is a simple pole; and if $m_G\!>\!2m$ then there is a higher-order pole.

The analysis of Dolan and Goddard~\cite{Dolan:2013isa} shows that if there is a pole in the 
integrand of $I_n$ at the point where the variables with indices in $T$ are equal, 
then $I_n$ will precisely pick up a propagator carrying the legs indexed by the numbers in $T$. 
To state this more precisely, we first have to introduce rescaled versions of the $S_i$:
\begin{align*}
S_i &= \frac{1}{\epsilon}[1+\mathcal{O}(\epsilon)]\tilde{S}_i\,, 
\hspace{5mm} \tilde{S}_i=\sum_{\substack{j \in T \\ j\neq i}}\frac{s_{ij}}{x_i-x_j} \,,
\hspace{5mm} &\text{for }i\in T\,, \\
S_i &= \hat{S}_i + \mathcal{O}(\epsilon)\,, \hspace{5mm} \hat{S_i}= \sum_{\substack{j 
\not \in T\\j\neq i}}\frac{s_{ij}}{z_i-z_j}+\sum_{j\in T}\frac{s_{ij}}{z_i}\,, &\text{for }i \not\in T\,.
\end{align*} 
The statement, then, is that if $\tilde{S}_i=0$ for all $i\in T\backslash \{n-m,n\}$, then
\begin{align*}
(x_{n-m}-x_{n})\tilde{S}_{n-m}=\tilde{S}_{n-m}=\frac{1}{2}\left(\sum_{i\in T} k_i\right)^2\,.
\end{align*}
When we use the global residue theorem, one of the residues contributing to $I_n$ will be the residue 
where $z_{n-m}$ is equal to zero and all $S_i$ except $S_{n-m}$ vanish for $3\!\leq\!i\!\leq\!n$. In the case where $m_G\!=\!2m$ and the 
pole is simple, the residue picks up a propagator and factorizes in two as follows:
\begin{align*}
&\hspace{-2.7cm}\text{Res}(S_3,S_4,\ldots,S_{n-m-1},z_{n-m},S_{n-m+1},\ldots,S_{n-1})=\\
\frac{1}{s_{n-m,n-m+1,\ldots,n}}&\oint\displaylimits_{\hat{S}_1=\cdots=\hat{S}_{n-m-1}=0} \hspace{-0.8cm}dz_3dz_4\cdots dz_{n-m-1} 
\frac{1}{\hat{G}(z_3,z_4,\ldots,z_{n-m-1})}\frac{1}{\prod_{i=3}^{n-m-1} \hat{S}_{i}}~\times\\
&\hspace{-2.5cm}\oint\displaylimits_{\tilde{S}_{n-m+1}=\cdots=\tilde{S}_{n-1}=0} \hspace{-0.8cm} dx_{n-m+1}dx_{n-m+2}\cdots dx_{n-1} 
\frac{1}{\tilde{G}(x_{n-m+1},x_{n-m+2},\ldots,x_{n-1})}\frac{1}{\prod_{i=n-m+1}^{n-1} \tilde{S}_{i}}\,,
\end{align*}
where
\begin{align*}
\lim_{\epsilon\rightarrow 0}\frac{G(z_3,z_4,\ldots,z_{n-1})}{\epsilon^{m_G}}=\hat{G}(z_3,z_4,\ldots,z_{n-m-1})
\tilde{G}(x_{n-m+1},x_{n-m+2},\ldots,x_{n-1})\,.
\end{align*}
From here, one could apply the counting rule to 
$$
\hat{G}(z_3,z_4,\ldots,z_{n-m-1})\, ~~~~~ \hbox{ \rm and}~~~~~\tilde{G}(x_{n-m+1},x_{n-m+2},\ldots,x_{n-1})\,,
$$ 
to look for further poles and iterate this procedure until the integrations have been completely 
carried out. Because the rescaled variables separate entirely from the non-rescaled ones, 
it is clear that the sets of external legs carried by the propagators will be either nested or disjoint. 
It also clear that in order for the residue to be non-zero, the number of poles encountered in this iterative procedure must equal the 
number of integrations---that is, $n-3$.

The above counting rule and factorization property are clearly specific to the chosen gauge. The power counting 
was particularly easy to perform because one of the variables $z_n$ in $T$ was gauge-fixed to zero. 
However, as $I_n$ is independent of the gauge choice, we need not concern ourselves with this issue. We also note that $I_n$ does not depend on the 
ordering of the external legs, as can be seen from the way it was originally expressed in the 
form prior to gauge-fixing. We are therefore free to relabel the legs and repeat the above argument. 
In fact, the elements of $T$ need not be consecutive: the counting rule still holds. Also, rather 
than applying  the counting rule to $G(z_3,z_4,\ldots,z_{n-1})$, we may apply the counting rule directly to the 
non-gauge-fixed factor $H(z)$. 

A minor comment: by gauge-fixing $z_1$ to infinity, we seem to have excluded the 
possibility of obtaining any propagator carrying external leg number one. This of course cannot
be true (and the integral is independent of the chosen gauge-fixing); and indeed, because of overall 
energy-momentum conservation, any propagator of a set of external legs equals the propagator 
carrying all the other external legs. In general, M\"obius invariance ensures that applying the counting rule to $H(z)$ on a subset of 
indices $\{1,2,\ldots,n\}$ is equivalent to applying the rule on the complement of the subset. 

To summarize, we have derived the following rules for evaluating $I_n$: 
\begin{itemize}
\item If there exists any subset $T\!\subset\!\{1,\ldots,n\}$ with $(m+1)$ elements such that $H(z)$ contains 
more than $2m$ factors of $(z_i-z_j)^{-1}$ with $i,j\!\in\!T$, then the integrand of $I_n$ has a higher-order 
pole, and $I_n$ cannot be written simply as a product of propagators. If, on the the other hand, the integrand of $I_n$ has no higher-order poles, 
we can evaluate $I_n$ through the following steps:
\item Find all poles by determining all subsets $T\!\subset\!\{1,\ldots,n\}$ such 
that $H(z)$ contains $2m$ factors of $(z_i-z_j)$ with $i,j\!\in\!T$. Assign to each pole a propagator 
of the form $1/\left(\sum_{i\in T}k_i\right)^2$. Complementary subsets are considered equivalent.
\item Whenever there are $(n-3)$ subsets that they are pairwise compatible, there is a residue which is equal to the product of the propagators of the subsets. 
Add together all the residues to obtain $I_n$.
\end{itemize}

As in the case of the string theory rules and their derivation, it is easy to extend these results to the case 
when $H(z)$ has a non-trivial numerator. The only change is that for every subset $T$ one should consider the following number: the number of factors $(z_i-z_j)^{-1}$ with 
$i,j\!\in\!T$ in $H(z)$ \emph{minus 
the number of factors}
$(z_k-z_l)$ \emph{with} $k,l\!\in\!T$ \emph{in} $H(z)$. If this number is equal to $2m$ there is a simple pole, 
and if it is higher there is a higher-order pole.

We will return to the case of higher-order poles, and how to evaluate CHY integrals in those cases in \mbox{section~\ref{pfafsection}}.

\subsection*{From String Theory to CHY via Two-Cycles} \label{dcsection}

The integration rules take on a particularly simple form when $H(z)$ can be written as a product of 
two ``cycles''---that is, when $H(z)\!=\!\text{Cycle}_a(z)\,\text{Cycle}_b(z)$ where the cycles are of the form
\begin{align}
\prod_{i=1}^n\frac{1}{z_{\sigma(i)}-z_{\sigma(i+1)}}\,,
\end{align}
with $\sigma\!\in\!\mathfrak{S}_n$, the permutation group on $n$ elements. 
In this case we may without loss of generality take one of these cycles to define the ordering of legs so that,
\begin{align}
\text{Cycle}_b(z)=\prod_{i=1}^n\frac{1}{z_i-z_{i+1}}\,,
\end{align}
and in applying the integration rules, one can instead consider subsets of consecutive numbers 
$T\!=\!\{j,j+1,\ldots,j+m\}$ such that Cycle$_b(z)$ contains $m$ factors of $(z_l-z_k)^{-1}$ with $l,k\!\in\!T$. 
Then the integration rules become {\em identical} to those derived for string theory integration, 
and we arrive at the identity:
\eqs{
\hspace{-70pt}&
\int dz_1dz_2\cdots dz_n \frac{(z_r-z_s)^2(z_s-z_t)^2(z_t-z_r)^2}{dz_rdz_sdz_t}~ 
\frac{\text{Cycle}_a(z)}{\prod_{i=1}^n(z_i-z_{i+1})}\prod_{\substack{i=1 \\ i\neq r,s,t}}^n\delta\big(S_i\big)\hspace{-50pt}
\\ 
\hspace{-70pt}=&
\lim_{\alpha'\!\to 0}\,{\alpha'\hspace{1pt}}^{n-3}\,
\int\prod_{i=3}^{n-1}
dz_i\,(z_1 - z_2)(z_2-z_n)(z_n - z_1)\,\text{Cycle}_a(z)
\prod_{1\leq i< j\leq n} |z_i-z_j|^{\alpha' s_{ij}}\,.\!\hspace{-50pt}}
In this way we derive explicitly the translation prescription (\ref{CHYstringlink}) that
was noted already in \mbox{ref.\ \cite{Bjerrum-Bohr:2014qwa}} and which links string theory integrands
to CHY integrands. We note that the appearance of $\text{Cycle}_b(z)$ in the CHY integrand
is what replaces the ordered integrations in string theory. 

\section{Reductions of Higher-Order Poles via Pfaffian Identities} \label{pfafsection}

We now return to the important question of how to evaluate CHY-type integrals when the
poles are of higher order. We remind the reader that this case was not treated in \mbox{section \ref{GRT}}. First of all,
in string theory higher-order poles generically correspond to terms that diverge so strongly that
the $\alpha'\!\!\to\! 0$ limit cannot be taken without resorting to analytical continuation. That is
why the simple correspondence between string theory integrands and CHY integrands~\cite{Bjerrum-Bohr:2014qwa}
is valid only after such double (tachyonic) poles have been manifestly cancelled in the integrand by
means of integrations by parts. Correspondingly, the CHY rule for integration must be modified.
We can solve this problem by applying a series of identities among CHY integrals.

Diagrammatically, we can represent the CHY integrals
\eq{I_n=\int dz_1dz_2\cdots dz_n \frac{(z_r-z_s)^2(z_s-z_t)^2(z_t-z_r)^2}{dz_rdz_sdz_t}~ H(z)
\prod_{\substack{i=1 \\ i\neq r,s,t}}^n\delta\big(S_i\big)\,,}%
where $H(z)$ has no factors of $(z_i-z_j)$ in the numerator in terms of so-called {\em 4-regular graphs} 
with $n$ vertices numbered from 1 to $n$. 
For each factor of $(z_i-z_j)^{-1}$ in $H(z)$, we draw an edge connecting vertices $i$ and $j$.

Now, if there is any subset $T\!\subset\!\{1,\ldots,n\}$ of $(m+1)$ elements such that $H(z)$ contains 
$2m+k$ factors of $(z_i-z_j)^{-1}$ with $i,j\!\in\!T$, then $I_n$ has a pole of order $k+1$. Graphically, 
integrals with third-order poles are those whose CHY diagrams consist of two or more separate graphs 
with no edges between them ({\it e.g.}\ a diagram that has two vertices connected by four edges); 
the integrals with double poles can be separated in two graphs with no edges between them if one removes 
two edges ({\it e.g.}\ a diagram with a triple line). Diagrams with higher-order poles appear in the Yang-Mills 
and gravity amplitudes in the CHY formalism.

The CHY formula for the tree-level gluon amplitude is given by~\cite{Cachazo:2013hca}
\eq{A_n=\int\!\!\chyInt\,\,\frac{\Pfprime\Psi}{(z_1-z_2)\cdots(z_n-z_1)}\,\,,}
while the CHY formula for the tree-level graviton amplitude is given by
\eq{M_n=\int\!\!\chyInt\,\,\big(\Pfprime\Psi\big)^2\,,}
where $\Psi$ is the $2n\!\times\!2n$ anti-symmetric given by
\begin{align}
\Psi = 
\begin{pmatrix}
A & -C^T \\
C & B
\end{pmatrix}\,,
\end{align}
with $A$ being an anti-symmetric $n\times n$ matrix of the form
\begin{align*}
A_{ij} = \frac{k_i \!\cdot\!k_j}{z_i-z_j}\,,
\end{align*} 
and
\begin{align}
&B_{ij}=
\begin{cases}
    \displaystyle\frac{\epsilon_i\!\cdot\!\epsilon_j}{z_i-z_j}\,, & \text{if $i \neq j$}\,,\\
    \displaystyle 0, & \text{if $i = j$}\,,
\end{cases}
\hspace{5mm}
&C_{ij}=
\begin{cases}
    \displaystyle\frac{\epsilon_i\!\cdot\!k_j}{z_i-z_j}\,, & \text{if $i \neq j$}\,,\\
    -\displaystyle \sum_{l\neq i}\frac{\epsilon_i\!\cdot\!k_l}{z_i-z_l}\,, & \text{if $i = j$}\,,
\end{cases}
\end{align}
and by $\Pfprime\Psi$ we denote the \emph{reduced} Pfaffian of $\Psi$, which is defined as
\begin{align*}
\Pfprime\Psi\equiv \frac{(-1)^{i+j}}{z_i-z_j}\,\text{Pf}\Psi_{i,j}\,,
\end{align*}
where $1\!\leq\!i \!<\! j\!\leq\!n$, and by $\Psi_{i,j}$ we denote the submatrix of $\Psi$ obtained by removing rows and columns $i$ and $j$. When evaluated on the scattering equations, $\Pfprime\Psi$ is independent of the choice of $i$ and $j$.

Because of the factor of $(z_1-z_2)\cdots(z_n-z_1)$ in the denominator of the gluon integrand, 
one can at most encounter double poles when expanding out all the terms in the Yang-Mills formula. 
The diagrams will always be connected. But in the case of gravity, one encounters double poles as 
well as triple poles.

As mentioned, CHY integrals with higher order poles cannot be directly evaluated with our integration 
rules provided above. But it is possible to express these more complicated integrals in terms 
of integrals covered by our integration rules. A neat tool for this is provided by Pfaffian 
identities. The Pfaffian identities that are of use to us concern the matrix $A$ considered above: 
\begin{align}
&A_{ij}=
\begin{cases}
    \displaystyle\frac{k_i\!\cdot\!k_j}{z_i-z_j}\,, & \text{if $i \neq j$}\,;\\
    \displaystyle 0\,, & \text{if $i = j$}\,.
\end{cases}
\end{align}
However, here we use the additional matrix only as an auxiliary tool, a 'generating function' for
CHY integral identities.

\vspace{10pt}
\subsection*{Even Multiplicity Reductions}

When $A$ is evaluated on a solution to the scattering equations, the Pfaffian vanishes. For even $n$, 
this fact provides us already with a non-trivial identity relating $(n-1)!!$ CHY diagrams. 
But there is another Pfaffian identity that is more practical because it relates only $2(n-3)!!$ 
CHY diagrams. This identity is based on the invariance of the reduced Pfaffian:  
\eq{\Pfprime A = \frac{(-1)^{i+j}}{z_i-z_j}\,\text{Pf}A_{i,j}\,. \label{evenid}}
As demonstrated in \mbox{refs.\ \cite{Cachazo:2013gna,Cachazo:2013hca,Cachazo:2013iea}}, on the solutions to the 
scattering equations, the value of 
$\Pfprime A$ is independent of the choice of $i$ and $j$.

Diagrammatically, the invariance of the reduced Pfaffian can be interpreted as follows. If we draw a 3-regular 
graph with $n$ vertices (we will from now on call this graph the `template'), 
then we need to superimpose a 1-regular graph in order to have a graphical representation of a CHY integral. 
The Pfaffian of $A$ can be regarded as the sum (with appropriate sign) over all possible 1-regular graphs without 
closed loops given $n$ vertices, where each term is multiplied with a Mandelstam variable $s_{ij}$. And the reduced Pfaffian can be 
regarded as the sum (with sign) over all possible 1-regular graphs without loops given 
$n$ vertices and one fixed edge, where each term is multiplied with a Mandelstam variable for each 
non-fixed edge. 

Equation~\eqref{evenid} then tells us the following. If on our 3-regular graph we connect any two vertices 
$i$ and $j$ with an additional edge and sum (with sign) over all ways of connecting the $n-2$ 
remaining vertices to form a 4-regular graph while multiplying each term with the proper 
Mandelstam variables, the result (the sum of the corresponding CHY integrals) will be independent 
of the choice of $i$ and $j$. 

\newpage
As an example, consider the following template:
\eq{\fig{-36.375pt}{1}{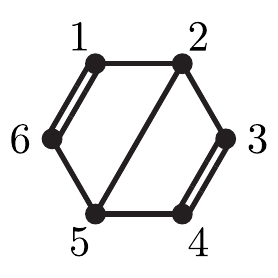}}
We note that this template is in itself not identifiable as a CHY integrand, as it is not
what is called a 4-regular graph: from each vertex must emanate precisely four lines. We will
now dress it up various different ways that turn this starting template into 4-regular graphs.

Using the fact that the reduced Pfaffian is the same whether we fix legs 1 and 4 or legs 4 and 5, 
we obtain the following identity:
\eqs{&s_{23}s_{56}\!\times\!\!\fig{-36.375pt}{1}{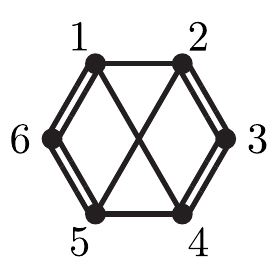}-s_{25}s_{36}\!\times\!\!
\fig{-36.375pt}{1}{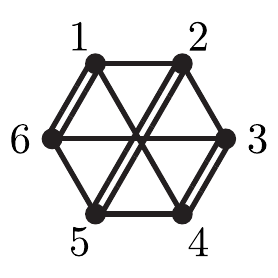}+s_{26}s_{35}\!\times\!\!\fig{-36.375pt}{1}{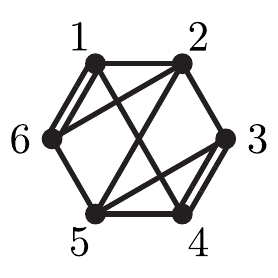}\\
\hspace{-12pt}=\;&s_{12}s_{36}\!\times\!\!\fig{-36.375pt}{1}{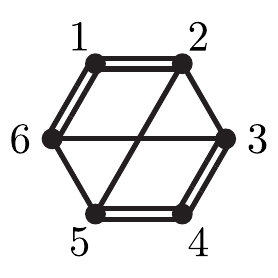}-s_{13}s_{26}\!
\times\!\!\fig{-36.375pt}{1}{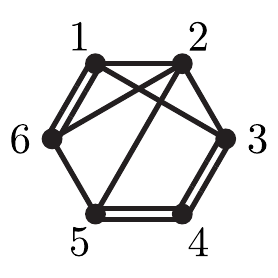}+s_{16}s_{23}\!\times\!\!\fig{-36.375pt}{1}{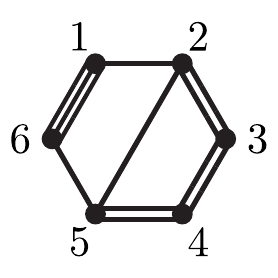}}
Read backwards, this identity rewrites a term with double pole (the last diagram) in terms
of diagrams with only single poles. These single-pole terms can be evaluated by means of our
integration rules. In practice, it can take a certain amount of experimentation before one
identifies the proper identity (or set of identities) that will completely rewrite a term
with a double pole in terms of single poles. The arbitrariness in this procedure is 
of the same kind as in string theory integrands, where also identifying a good choice of partial integrations
must rely on a certain amount of experimentation. 

\newpage
\subsection*{Odd Multiplicity Reductions}

When $n$ is odd, the Pfaffians of $A$ and $A_{i,j}$ vanish trivially and the identities 
described above are of no use. But if by $A_i$ we denote the sub-matrix of $A$ obtained by removing 
only row and column $i$ from $A$, then $\text{Pf} A_i=0$ for all $i$. 
To see this, one can, without loss of generality, consider $A_1$. 
If for each $j\!=\!1,\ldots,n-1$ one multiplies the $j^{\text{th}}$ row with $k_1\!\cdot\!k_{j+1}$, then their 
sum gives zero when evaluated on the scattering equations, implying that the rows of $A_1$ are not 
linearly independent.

The vanishing of $\text{Pf} A_i$ provides an identity that relates $(n-2)!!$ CHY diagrams. 
Graphically, the identity can be represented as follows:
As a template we draw a graph with $n$ vertices and $\frac{1}{2}(3n+1)$ edges such that there are 
three edges incident to each vertex except for one vertex to which four edges are incident. 
If we then sum (with sign) over all the ways of drawing the remaining $\frac{1}{2}(n-1)$ edges so 
that we obtain a 4-regular graph, while multiplying each term with the Mandelstam variable $s_{ij}$ 
if an edge connecting vertices $i$ and $j$ is added to the template, then the result will be zero.

As an example, consider the following template:
\eq{\fig{-36.375pt}{1}{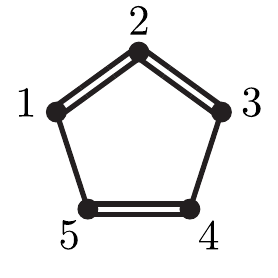}\,}
Only point 2 has the needed number of lines. Points 1, 3, 4 and 5 need additional connections 
in order to produce 4-regular graphs.
Performing the weighted sum over the different ways of completing the diagram, we get zero:
\eq{\hspace{-30pt}0=s_{34}s_{51}\!\fig{-36.375pt}{1}{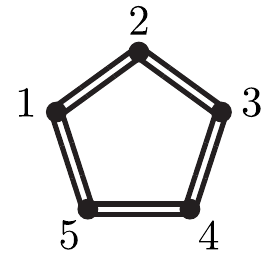}-s_{35}s_{14}\!
\fig{-36.375pt}{1}{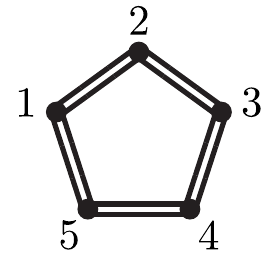}+s_{13}s_{45}\!\fig{-36.375pt}{1}{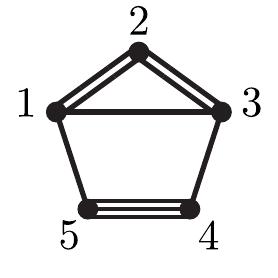}\,.}
Read backwards, this describes the double pole of the last diagram in terms of the two
other diagrams with only single poles. We have thus provided examples for both odd and even $n$
on how to evaluate CHY integrands with higher-order poles.

In some cases the use of such Pfaffian identities allows one to immediately express an unknown 
CHY diagram in terms of some that can be evaluated with our integration rules. But in general 
it will be necessary to invoke several Pfaffian identities involving several new and unknown 
CHY diagrams and then eventually solve the system of equations. A systematic procedure
is to start with simple diagrams and then work towards more complicated ones. A diagram with four 
edges connecting the same two vertices can only appear in Pfaffian identities where the template has 
two vertices connected with three edges, so the identities involving quadruple-line diagrams 
will always involve only triple- and quadruple-line diagrams. But these in turn can be re-expressed 
in terms of diagrams with fewer quadruple- and triple-lines by using Pfaffian identities. The 
process can be iterated until one reaches diagrams that can be written simply as sums of products 
of propagators. Again, there is a direct analogy between this procedure and that of solving
a set of integration-by-parts identities in string theory.

The identities below will serve as examples of how to reduce diagrams to simpler ones in 
the case of $n\!=\!6$:
\eqs{\hspace{-10pt}\fig{-36.375pt}{1}{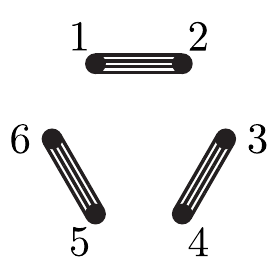}=&\phantom{-\,}\frac{s_{16}s_{23}}
{s_{34}s_{56}}\fig{-36.375pt}{1}{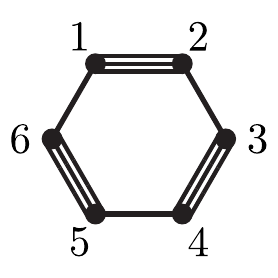}\hspace{20pt}+\hspace{-20pt}\fwboxR{70pt}
{\frac{s_{35}s_{46}}{s_{34}s_{56}}}\fig{-36.375pt}{1}{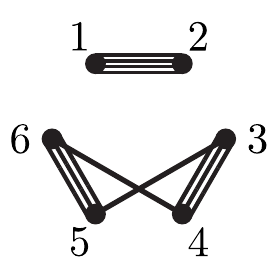}\\&{-\,}\frac{s_{13}s_{26}}
{s_{34}s_{56}}\fig{-36.375pt}{1}{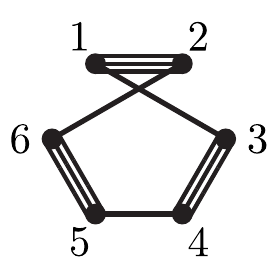}+\fwboxR{70pt}{\frac{s_{36}(s_{12}\mi\,s_{45})}
{s_{34}s_{56}}}\fig{-36.375pt}{1}{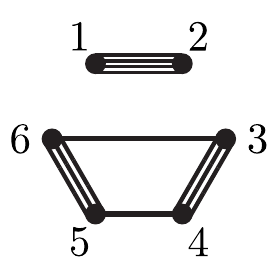}\,.}
 These four simpler diagrams can then be further reduced themselves. The first and third are related by a 
 reordering of the external points and so one need only decompose one of them, for example:
 \eqs{\hspace{-10pt}\fig{-36.375pt}{1}{6pt_template_3_2}=&\phantom{-\,}\frac{s_{35}s_{46}}{s_{34}
 s_{56}}\fig{-36.375pt}{1}{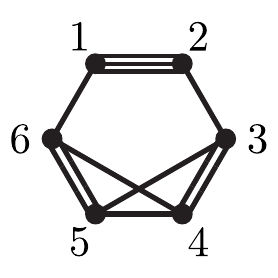}\hspace{20pt}
 +\hspace{-20pt}\fwboxR{70pt}{\frac{s_{15}s_{24}}
 {s_{34}s_{56}}}\fig{-36.375pt}{1}{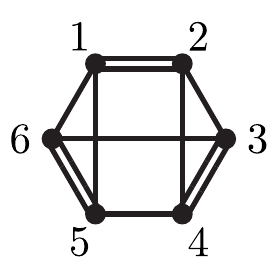}\\&{-\,}\frac{s_{14}s_{25}}{s_{34}s_{56}}
 \fig{-36.375pt}{1}{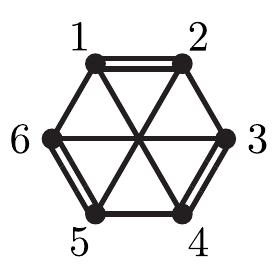}
 +\fwboxR{70pt}{\frac{s_{45}(s_{12}\mi\,s_{36})}{s_{34}s_{56}}}\fig{-36.375pt}{1}{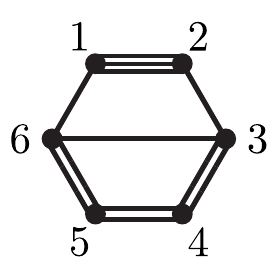}\,.}
Diagrams 2 and 3 on the right-hand side are now in a form so that we can apply the integration rules to them. The first and last 
have to be simplified further. The first one can be reduced thus:
\eqs{\hspace{-10pt}\fig{-36.375pt}{1}{6pt_template_4_1}=&\phantom{-\,}\frac{s_{15}s_{26}}{s_{12}
 s_{56}}\fig{-36.375pt}{1}{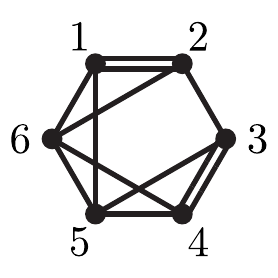}\hspace{20pt}+\hspace{-20pt}\fwboxR{70pt}{\frac{s_{13}
 s_{46}}{s_{12}s_{56}}}\fig{-36.375pt}{1}{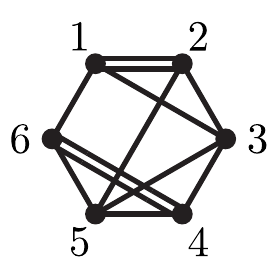}\\&{-\,}\frac{s_{14}s_{36}}{s_{12}s_{56}}
 \fig{-36.375pt}{1}{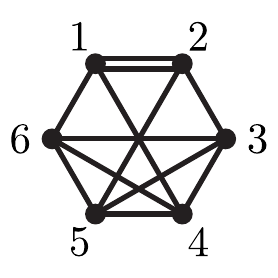}+\fwboxR{70pt}{\frac{s_{16}(s_{34}\mi\,s_{25})}{s_{12}s_{56}}}
 \fig{-36.375pt}{1}{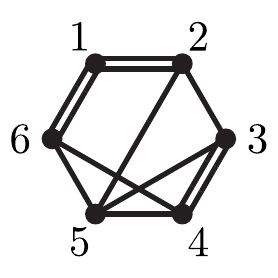}\,.}
We hope these examples suffice to illustrate the general method.

For simplicity, we have in this section considered the special case when $H(z)$ has a trivial numerator. 
But integrals whose integrands have non-trivial numerators can also be re-expressed via Pfaffian identities. 
Such integrals can be represented diagrammatically by adding to the type of diagram described above a dotted 
line connecting vertices $i$ and $j$ for every factor of $(z_i-z_j)$ in the numerator. In that case M\"obius 
invariance dictates that the number of normal lines minus the sum of dotted lines incident on each vertex 
should equal four.  And in a manner completely identical to the above, one can construct templates for 
diagrams that have dotted lines and apply Pfaffian identities to the templates.

It is necessary to calculate CHY integrals that also have cross ratios if one wishes to calculate 
gluon or graviton amplitudes in the CHY formalism by expanding out $\Pfprime\Psi$. This is because of the diagonal 
entries of the matrix $C$:
\begin{align}
C_{l l}=-\sum_{i\neq l}\frac{\epsilon_l\!\cdot\!k_i}{z_i-z_l}\,,
\end{align}
always (because of momentum conservation) can be expanded in a manifestly M\"obius invariant form where, after using momentum conservation, it reads, for example
\begin{align}
C_{l l}=\sum_{i\neq l}^{n-1}\frac{\epsilon_l\!\cdot\!k_i ~(z_i-z_n)}{(z_i-z_l)(z_n-z_l)}\,.
\end{align}
Also such integrals with cross ratios can be calculated using our rules, see below.

It should be noted that there is a class of diagrams that cannot be reduced by using identities concerning 
$\text{Pf} A$. These are diagrams for which there is 
no template. An example is this one,
\eq{\fig{-36.375pt}{1}{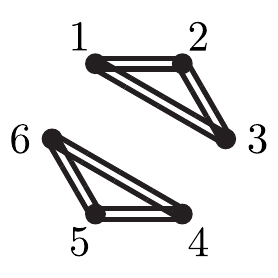}}
Such diagrams do not appear when expanding out the CHY formula for gluon amplitudes, 
but they do appear in the graviton case. To tackle them, one can consider Pfaffian identities 
involving the full matrix $\Psi$ rather than just $A$. It would be interesting to systematically
explore the space of CHY identities in this way.

\section{Comparison to other CHY Integration Methods}
The possibility of calculating CHY integrals without actually solving the scattering equations has been explored
in the literature previously. As an alternative to our integration rules Kalousios was able to exhaustively 
work out the $n=5$ case using the Vieta formul\ae\ that relate sums of roots of polynomials to their 
coefficients~\cite{Kalousios:2015fya}. 
Cachazo and Gomez~\cite{Cachazo:2015nwa} have provided an exhaustive 
treatment of the six-point case but the method does involve several non-trivial graph theoretical considerations which can be avoided
invoking Pfaffian identities. Also, their basic integrals constitute a smaller 
class than those that can be evaluated with our integration rules.
Their basic integrals are those, considered in section~\ref{dcsection}, where $H(z)$ can be written as 
the product of two-cycles: $H(z) =\text{Cycle}_a(z)\,\text{Cycle}_b(z)$. As shown by CHY in 
\mbox{ref.\ \cite{Cachazo:2013iea}}, such integrals evaluate to the sum of all Feynman diagrams that 
are compatible with the orderings of both the cycles. But the class of CHY integrals that evaluate 
to a sum of Feynman diagrams, {\it viz.}\ the class of CHY integrals that can be evaluated with our 
integration rules, is larger. As an example, we can consider CHY integrals whose integrands 
cannot be decomposed into two cycles.

To illustrate this last point, consider the following
CHY diagram which cannot be decomposed into two cycles:
\eq{\fig{-36.375pt}{1}{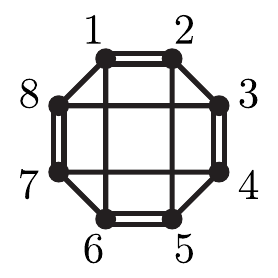}=\frac{1}{s_{12}s_{34}s_{56}s_{78}s_{1256}}\,.}
Nevertheless, we easily evaluate it to the result shown with the integration rules by noting that the diagram 
has a total of five subsets $T$ of vertices (excluding their complements) with enough edges connecting 
them to produce a propagator: $\{1,2\}$, $\{3,4\}$, $\{5,6\}$, and $\{7,8\}$, which all contain two vertices 
connected by two edges, and $\{1,2,5,6\}$, which contains four vertices connected by six edges. 
And these five subsets are all either nested or disjoint.

Another example of a CHY diagram that cannot be decomposed into two cycles but which can be easily evaluated using our integration rules, is the following:
\eq{\fig{-56.375pt}{1}{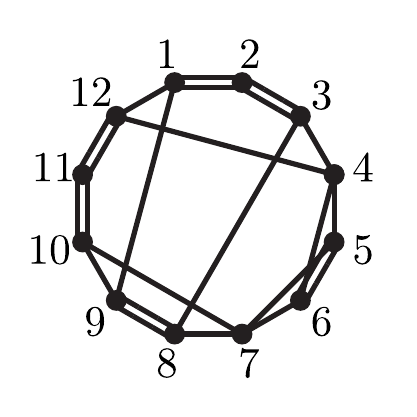}}
To evaluate it, we
start by listing the subsets of vertices with (2 times the number of vertices minus 2) edges connecting them:
\begin{align}
&\begin{rcases}
\{1,2\}\\
\{2,3\}\\
\{5,6\}\\
\{8,9\}\\
\{10,11\}\\
\{11,12\}&
\end{rcases}
\text{two vertices, two edges}
\\
&\begin{rcases}
\{1,2,3\}\\
\{4,5,6\}\\
\{5,6,7\}\\
\{10,11,12\}&
\end{rcases}
\text{three vertices, four edges}
\\
&\begin{rcases}
\{4,5,6,7\}&
\end{rcases}
\text{four vertices, six edges}
\\
&\begin{rcases}
\{1,2,3,8,9\}&
\end{rcases}
\text{five vertices, eight edges}
\end{align}
All these sets are compatible with each other in the sense that any two sets are either nested or 
disjoint---except for the following three overlapping sets:
\begin{align*}
&\{1,2\}\text{ overlaps with }\{2,3\}
\\
&\{10,11\}\text{ overlaps with }\{11,12\}\text{, and}
\\
&\{4,5,6\}\text{ overlaps with }\{5,6,7\}.
\end{align*}
This leaves $2^3$ different ways of combining 9 compatible subsets. Summing over the 
corresponding products of propagators we get the final result:
\begin{align}
\left(\frac{1}{s_{23}}+\frac{1}{s_{12}}\right)\frac{1}{s_{56}}\frac{1}{s_{89}}\left(\frac{1}{s_{10\,11}}
+\frac{1}{s_{11\,12}}\right)\frac{1}{s_{123}}
\left(\frac{1}{s_{456}}+\frac{1}{s_{567}}\right)\frac{1}{s_{10\,11\,12}}\frac{1}{s_{4567}}\frac{1}{s_{12389}}\,.
\end{align}
As mentioned, there are also cases where the integration rules can be applied to integrands 
with non-trivial numerators---provided, as before, that the integral has no higher order poles. 
Because the presence of factors in the numerator forces the denominator to have more factors 
than otherwise in order for the integral to retain M\"obius invariance, integrands with numerators 
will in most cases have higher order poles. But those that do not can readily be evaluated with the 
integration rules of this paper. One merely counts dotted lines as negative solid lines; they carry negative 
weight. 

To illustrate, consider a CHY integral with
\begin{align}
H(z) = \frac{(z_2\!-\!z_6)}{(z_1\!-\!z_2)^2(z_1\!-\!z_6)^2
(z_2\!-\!z_3)^2(z_2\!-\!z_4)(z_3\!-\!z_4)(z_3\!-\!z_5)(z_4\!-\!z_5)(z_4\!-\!z_6)(z_5\!-\!z_6)^2}\,,
\end{align}
which can be represented by the following diagram:
\eq{\fig{-36.375pt}{1}{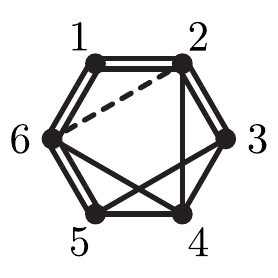}}
To evaluate the integral we first enumerate the subsets of points with net (2 times the number of points minus 2) lines connecting them:
\begin{align}
&\begin{rcases}
\{1,2\}\\
\{1,6\}\\
\{2,3\}\\
\{5,6\}&
\end{rcases}
\text{two points, two lines}
\\
&\begin{rcases}
\{1,2,3\}\\
\{2,3,4\}&
\end{rcases}
\text{three vertices, four lines}
\end{align}
The points $\{1,2,6\}$ are connected by four normal lines, but the dotted line counts minus one, so this 
subset does not make it to the list. Of the subsets that are on the list, we can form the following four 
maximal groups of compatible subsets:
\begin{align}
& \{1,2\},\{5,6\},\{1,2,3\} \\
& \{1,6\},\{2,3\},\{2,3,4\} \\
& \{2,3\},\{5,6\},\{1,2,3\} \\
& \{2,3\},\{5,6\},\{2,3,4\}  
\end{align}
We conclude that the integral is given by:
\begin{align}
\left(\frac{1}{s_{12}}+\frac{1}{s_{23}}\right)\frac{1}{s_{56}s_{123}}+
\left(\frac{1}{s_{16}}+\frac{1}{s_{56}}\right)\frac{1}{s_{23}s_{234}}\,.
\end{align}
This shows how straightforward it is to apply our rules to integrands with
non-trivial numerators.

\section{Including a Pfaffian: Integration Rules for $\phi^4$-theory}
Recently, Cachazo, He and Yuan~\cite{Cachazo:2014xea} have demonstrated how $\phi^4$-theory
can be treated in the scattering equation formalism. Using dimensional reduction of Yang-Mills theory
with the compactified gauge bosons taking on the role of scalars, they arrive at an integral
that generically looks as follows for $n$ even:
\begin{align}
\mathcal{I}_n=\int dz_1dz_2\cdots dz_n &\frac{(z_r-z_s)^2(z_s-z_t)^2(z_t-z_r)^2}{dz_rdz_sdz_t}
\frac{\Pfprime A}{C(z)}\prod_{i=1}^n\frac{1}{z_i-z_{i+1}}\prod_{\substack{i=1 \\ i\neq r,s,t}}^n\delta\big(S_i\big)\,, \label{CHYintPfaf}
\end{align} 
where $C(z)$ is a product of differences $(z_i-z_j)$ with each $z_i$ appearing in exactly one factor, 
and $A$ is as defined in the previous section, where we also introduced the reduced Pfaffian of
this matrix. A crucial property of $\mathcal{I}_n$ is that it does not depend on which rows and columns are removed from $A$ in 
calculating the reduced Pfaffian.

We now consider the special case when $C(z)$ is a ``connected perfect matching''~\cite{Cachazo:2014xea}.
It then has the property that if one selects any proper subset $T\!\subset\!\{1,2,\ldots,n\}$ of 
consecutive numbers, $C(z)$ will contain at least one factor $(z_k-z_l)$ with $k\!\in\!T$ and 
$l\!\not\in\!T$. As shown in \mbox{ref.\ \cite{Cachazo:2014nsa}}, in this special case it happens, miraculously, 
that when $C(z)$ can be represented graphically as a tree-level $\phi^4$ Feynman diagram, $\mathcal{I}_N$ evaluates 
to just that Feynman diagram, while $\mathcal{I}_n=0$ otherwise. 
This unusual behavior of $\mathcal{I}_n$  can also be phrased in a way that is highly reminiscent of the 
above integration rules. It seems worthwhile to provide this additional integration rule.

All expressions $\mathcal{I}_n$ for which $C(z)$ is a perfect matching of the above form can be 
evaluated by the following procedure:
\begin{itemize}
\item Consider in turn all proper subsets $T\!\subset\!\{2,3,\cdots,n\}$ of $m_T$ consecutive numbers (1 and $n$ are to be considered consecutive) so that
$m_T$ is odd and bigger than 1, {\it i.e.}, $m_T\!\in\!\{ 3,5,7,\ldots,n-3\}$. Complementary subsets are considered equivalent.
\item For each of these subsets $T$, count the number of factors $(z_i-z_j)$ in $C(z)$ for which 
$i\!\in\!T$ and $j\!\in\!T$. We denote this the number of {\em connections} for the given subset.
\item If a subset $T$ has $(m_T-1)/2$ connections, we shall say that it is fully connected. 
With each fully connected subset $T\!=\!\{i,i+1,i+2,\ldots,i+m_T-1\}$ associate a propagator $1/s_{i,i+1,i+2,\ldots,i+m_T-1}$.
\item Count the number of fully connected subsets. If it is less than $(n-4)/2$, then $I_n=0$. 
If the number is equal to $(n-4)/2$, then $I_n$ is given by the product of the propagators of the 
fully connected subsets.
\end{itemize}
Interestingly,
these integration rules can also be viewed in the light of the correspondence between string theory
and CHY-type integrals. Indeed, upon compactification of the open {\em bosonic} string in precisely the same
manner as in \mbox{ref.\ \cite{Cachazo:2014xea}} one gets
\begin{align}
\mathcal{I}_n = \lim_{\alpha'\!\to 0}\,{\alpha'}^{(n-4)/2}\,
\int\prod_{i=3}^{n-1}
dz_i\,{(z_1 - z_{2})(z_{2}-z_n)(z_n - z_1)\over C(z)^2}
\prod_{1\leq i< j\leq n} |z_i-z_j|^{\alpha' s_{ij}}\!
\,,\end{align}
This identity between the bosonic string theory and CHY expressions holds even in the cases where $C(z)$ is 
not a connected perfect matching. For those cases we do not have direct integration rules because they
correspond to higher-order poles. It is an interesting fact that the result of CHY integration 
exactly matches the string theory computation after performing analytic continuation to the region
around $\alpha'\!\!=\!0$.

An example of such a disconnected perfect matching is 
\begin{align}
C(z)=(z_1-z_2)(z_3-z_6)(z_4-z_8)(z_5-z_7)\,, \label{disconC}
\end{align}
which can be represented diagrammatically as
\eq{\fig{-36.375pt}{1}{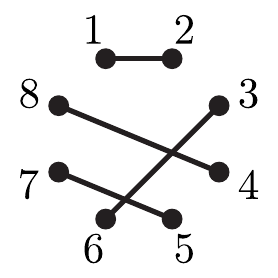}}
In the $z_1=\infty$, $z_2=1$, $z_8=0$ gauge, the string theory integral takes on the form
\begin{align}
\lim_{\alpha'\!\to 0}(\alpha')^2\prod_{i=3}^{7}\left(\int_0^{z_{i-1}}dz_i\right)
\frac{1}{(z_3-z_6)^2z_4^2(z_5-z_7)^2}\prod_{i=2}^{n-1}\prod_{j=i+1}^n(z_i - z_j)^{\alpha's_{ij}} \,.
\end{align}
In carrying out the integration it can be convenient to change to rescaled variables $x_i\!\equiv\!z_i/z_{i-1}$. 
In terms of the new variables, the divergent integration regions that contribute a factor of $(\alpha')^{-1}$ are: 1. 
the region where $x_6$ and $x_7$ tend to one; 2. the region where $x_4$, $x_5$, $x_6$, and $x_7$ tend to 
one; and 3. the region where $x_4$ tend to zero. Of these three regions, 2 and 3 are incompatible, though 
they are both compatible with region 1. After carrying out the integrations over $x_4$ to $x_7$ we are therefore left with
\begin{align}
\lim_{\alpha'\!\to 0}\int_0^1dx_3\frac{x_3^{\alpha's_{345678}}}{x_3^2}\left(\frac{(1-x_3)^{\alpha's_{23}}}{s_{45678}}+\frac{(1-x_3)^{\alpha'(s_{23}+s_{24}+s_{25}+s_{26}+s_{27})}}{s_{34567}}\right)\frac{1}{s_{567}}\,.
\end{align}
Interpreted as a Riemann integral, it is not analytic in the vicinity of $\alpha'\!\!=\!0$. But after analytical 
continuation the integral is finite and will be seen, after expanding the beta function, to yield a value of
\begin{align}
\left(1+\frac{s_{23}}{s_{345678}}\right)\frac{1}{s_{45678}}\frac{1}{s_{567}}+\left(1+\frac{s_{23}+s_{24}+s_{25}+
s_{26}+s_{27}}{s_{345678}}\right)\frac{1}{s_{34567}}\frac{1}{s_{567}}\,.
\end{align}
This matches precisely the result obtained by performing the CHY integral~\eqref{CHYintPfaf} with $C(z)$ as given 
in~\eqref{disconC}. This is easily seen by expanding out $\Pfprime A$ and employing our integration rules.

\section{Conclusions}

To summarize, we have provided a set of integration rules for both string theory
integrands and CHY integrands, and demonstrated the equivalence between the two
in the $\alpha'\!\!\to\! 0$ limit. This puts on a firm footing the transcription between
superstring theory integrands and CHY integrands that was proposed in \mbox{ref.\ \cite{Bjerrum-Bohr:2014qwa}}. In the string theory integral the crucial ingredient
was a systematic description of the terms that provide the precise divergence
as $\alpha'\!\!\to\! 0$ needed to recover a finite overall result in that limit.
In the CHY integral, the analogous tool comes from the global residue theorem,
which picks up all contributions to the CHY integral.

The main advantage of our CHY integration rules is that they are simple, algorithmic
and easy to program. To solve CHY integrals one does not need to explicitly find
all $(n-3)!$ solutions of the scattering equations, but can rather apply these
rules. The result is identical.

In the process, we hope to have shed more light on the correspondence between
string theory integrals and CHY integrals. In particular, we have derived quite
simple algorithmic rules for how to evaluate the field theory limit of string theory
amplitudes. We have also shown how the global residue theorem allows us to derive
the precise analog in the CHY formalism. 

There are various outgrowths of these results that could be interesting to investigate.
First of all, it would be of interest to elucidate the precise connection between our integration
rules and those of \mbox{ref.\ \cite{Cachazo:2015nwa}}. It should also be possible to use
the global residue theorem to explicitly derive CHY integration rules for higher-order
poles, rather than, as done here, reduce those integrals by means of Pfaffian
identities.

Knowing integration rules, one can work backwards and try to construct CHY integrands
for other theories. This might possibly shed light on how to define $\phi^p$-theory with $p$ different from
3 or 4. It could also be interesting to find a more direct link between CHY 
integrals and individual Feynman diagrams in a given field theory. We hope to
return to some of these issues in a future publication.

\vspace{10pt}
\subsection*{Acknowledgements}
This work has been supported in part by a MOBILEX research grant from the Danish Council for 
Independent Research (JLB).

\newpage
\providecommand{\href}[2]{#2}\begingroup\raggedright\endgroup

\end{document}